\documentclass[a4paper,11pt]{article}
\pdfoutput=1 
\usepackage{jinstpub} 

\title{\boldmath Electromagnetic Data Libraries: recent evolutions and new perspectives}

\author[a]{D. C. Duma,}
\author[b]{S. Parlati,}
\author[c,1]{M. G.  Pia,\note{Corresponding author.}}
\author[a]{E. Ronchieri}
\author[c]{and P. Saracco}

\affiliation[a]{INFN CNAF,\\some-street, Italy}
\affiliation[b]{ INFN Laboratori Nazionali del Gran Sasso, \\Via Acitelli 22, 67100 L'Aquila, Italy}
\affiliation[c]{INFN Sezione di Genova, \\Via Dodecaneso 33, 16146 Genova, Italy}

\emailAdd{mariagrazia.pia@ge.infn.it}

\abstract{
This paper summarizes the current status of the electromagnetic data libraries,
reviews recent experimental validation results, highlights open issues and
introduces new perspectives for the future of these data libraries taking shape
in the context of INFN research. Special emphasis is given to the
characteristics of reliability, transparency and openness, along with
opportunities for the improvement and the extension of the physics content.}


\proceeding{15$^{\text{th}}$ Topical Seminar on Innovative Particle and Radiation Detectors \\
  14-17 October 2019\\
  Siena, Italy}

\begin{document}
\maketitle
\flushbottom

\section{Introduction}
\label{sec:intro}

Evaluated data libraries are tabulations of physics quantities (cross sections,
nuclear and atomic parameters, secondary particle spectra etc.), which incorporate
the body of knowledge of theoretical computations, experimental measurements or
both in the physics area they pertain to.
Their intended purpose is to represent the state of the art in the field,
assessed and assembled by experts, and made available to  the community 
for a wide variety of scientific, engineering and industrial uses.

Evaluated data libraries are an essential instrument in many 
computational systems.
They are an essential component for Monte Carlo particle transport and detector
research, since they are the basis for reliable modelling and fast computation of
particle interactions with matter.

Various nuclear data libraries are available, such as 
BROND (Russian Evaluated Neutron Data Library) \cite{BROND:2016}, 
CENDL (Chinese Evaluated Nuclear Data Library) \cite{cendl31}, 
ENDF/B (Evaluated Nuclear Data File) \cite{ENDF-VIII.0}, ENSDF \cite{ENSDF},
JEFF (Joint Evaluated Fission and Fusion File) \cite{Koning2006},  
JENDL (Japanese Evaluated Nuclear Data Library) \cite{JENDL3.3}
and TENDL \cite{tendl}.
A collection of three evaluated data libraries (EADL \cite{eadl}, EEDL \cite{eedl}, EPDL \cite{epdl97}), 
concerning electron and photon interactions and atomic parameters, has
represented the reference for electromagnetic interactions for several decades.

This paper briefly summarizes the current status of the electromagnetic data libraries,
reviews recent experimental validation results, highlights open issues and
introduces new perspectives for the future of these data libraries taking shape
in the context of INFN research.
Special emphasis is given to the characteristics of reliability, transparency
and openness, along with opportunities for the improvement and the extension of
the physics content.

\section{Overview of electromagnetic data libraries}

The Evaluated Atomic Data Library (EADL), which encompasses  atomic
parameters and atomic relaxation data, the Evaluated Electron Data Library (EEDL) and the
Evaluated Photon Data Library (EPDL),  which collect cross section
data and related physical quantities pertinent to electron and photon
interactions with atoms, are independently released within ENDF/B and by the
IAEA (International Atomic Energy Agency), which collctively identifies them as EPICS (Electron Photon Interaction Cross Sections) \cite{EPICS2014}.
The data in EEDL and EPDL are the outcome of theoretical calculations;
those in EADL are partly empirical and partly theoretical.

These data libraries are used by major Monte Carlo codes, such as EGS \cite{egs5}
\cite{egsnrc}, FLUKA \cite{fluka1} \cite{fluka2}, Geant4 \cite{g4_nim}
\cite{g4_tns} \cite{g4_nim2}, ITS \cite{its2018}, MCNP \cite{mcnp6} and Penelope
\cite{penelope2014}.


The latest versions of these data libraries at the time of writing this paper
(\cite{eadl2017}, \cite{eedl2017}, \cite{epdl2017}) are identified as EPICS 2017.
From a physics perspective, the major modification with respect to the previous
versions concerns the atomic binding energies, which affect other depending data
in the electron and photon data libraries.
The compilations of binding energies used to produce the new data libraries were
identified as more accurate than the previous EADL values in a previously
published validation test \cite{tns_binding}.

Nevertheless, inconsistencies are observed in the propagation of the modified
binding energies to the tabulations of dependent quantities, which could be the
source of disruptive effects in Monte Carlo simulations.
These and other inconsistencies were observed in the EPICS 2017 release \cite{tns_epics};
concerns are also raised by the lack of proper configuration management and version control, which hinders 
the reproducibility of simulation results based on the data libraries.

Although these data libraries play a critical role in many areas of experimental
physics research, they do not cover all the needs of current and future
experiments.
They are affected by limitations both at low ($<1$ keV) and high energies, which
they nominally cover up to 100~GeV.
Relevant quantities for particle transport (e.g. stopping powers of
charged particles) are not included.
The theoretical calculations on which they are based are outdated in some cases,
where more recent developments have been identified by validation tests as 
more accurate.
A major shortcoming  is the lack of any estimate of their 
associated uncertainties, which hinders uncertainty quantification and sensitivity analysis
in the experimental contexts where they are used.



%

\section{New perspectives for physics data libraries}

A wide international discussion is in progress to identify areas where
improvements would be beneficial to the future of physics data libraries and,
consequently, to the multidisciplinary community that relies on them for scientific research
and engineering applications.

Although functionality is obviously a prime concern \cite{bernstein_2019}, other critical aspects are
equally important to ensure the long term fruition, maintainability and reliability
of physics data libraries.
They are the object of extensive investigation to address problematic areas with 
common solutions, beyond the borders of single data libraries and the
specificity of their functionality.

\subsection{Extension of physics functionality}

Validation tests have identified areas where the current EEDL and EPDL content
no longer represents the state of the art and theoretical approaches 
other than those originally adopted in the compilation of these data libraries 
exhibit better compatibility with experimental measurements:
for instance, photon elastic scattering differential cross sections
\cite{tns_rayleigh} and the total cross sections for electron impact ionization
\cite{tns_beb}.

Nevertheless, it is worthwhile to note that more modern calculations are not
necessarily more accurate than EEDL and EPDL tabulations: this has been observed
in cross sections for electron impact ionization \cite{tns_bebshell} and in
photoelectric cross sections \cite{tns_photoel}.


\subsection{Physics content and data preservation}

The physical content of data libraries is the focus of interest,
since it is required to fulfil the needs of current and future use in diverse areas.
Coverage of low and high energy particle interactions is a critical issue for
many experimental scenarios; limitations on both ends, which are present in  the
the data libraries commonly  used in particle transport, should be addressed by 
theoretical and experimental efforts to extend the current capabilities.

Physics data libraries collect a vast body of knowledge from theoretical and
experimental sources, evaluators and computational physics experts.
Plans for data preservation, which are currently focused on experimental data
\cite{Kogler_2012}, should be extended to ensure continued access also to the
content of physics data libraries, as well as to the associated computational
environment, such as the theoretical codes that produce the tabulated data and
processing codes that make them usable.
Data preservation requires setting up a series of managed activities to fulfil 
this purpose.
Ideally, a preservation program should care for the expertise associated with
the data.

\subsection{Openness and transparency}

Major physics data libraries, such as ENDF/B \cite{ENDF-VIII.0}, JEFF
\cite{Koning2006} and JENDL \cite{JENDL3.3}, are openly available.
Other data sets of wide interest are available from government and academic
centres: for instance, spectroscopic and dosimetry data sets are distributed by
NIST through its Physical Reference Data service \cite{NISTphysref}.
Nevertheless, their permanent availability cannot take for granted, if they are subject 
to the fluctuations a single national control or of funding constraints.

Other important data collections are published in journals or reports,
which are not generally openly accessible and may be subject to copyright restrictions.
Notable examples are the compilations of charged particle stopping powers
published by the International Commission on Radiation Units and Measurements (ICRU)
\cite{icru37}, \cite{icru49}, \cite{icru73}, \cite{icru73errata}.
Some data tabulations, e.g. the cross section calculations described in
\cite{sabbatucci_2016}, are only privately available from their authors.

Some physics data compilations are distributed by private web sites, or embedded
in commercially available reports, or are associated with the Monte Carlo codes that use
them, which may be subject to restrictions regarding their distribution.

Often, although the libraries may be openly available, the origin of their
content is incompletely documented.
Consistent use of the data in experimental applications requires their
traceability \cite{cleland_2014}: it means that methods, assumptions, approximations, parameters
etc. involved in producing the data, and their interrelations should be
documented.

Traceability ensures the reproducibility of the data, which is necessary for their
maintainability. 
It is supported by rigorous configuration and change management \cite{haug2001managing} in the 
production of the data; with respect to their use, it requires version
control of the released data.
Substantial improvements are required in this area: for instance, the
distribution of atomic data libraries, used by most major Monte Carlo transport
systems, is severely deficient regarding version control \cite{tns_epics}.

Besides openness, transparency is a key requirement for physics data.
Along with transparency of the production process of data libraries,
transparency of their performance is required for using them properly: limits of their
usability and uncertainties of the data should be openly accessible.
While substantial progress has been made in the evaluation of uncertainties 
associated with nuclear data,  estimates of the uncertainties of
atomic data are currently missing.
Filling this gap is far from trivial, since it requires the elaboration of a
generally accepted methodology to associate uncertainties with the tabulated theoretical 
calculations.


A related issue is the openness of the theoretical codes that calculate the 
physics data tabulated in the libraries.
Open source availability of these codes, along with the documentation of the underlying assumptions and
approximations embedded in the calculations, would be highly valuable to ensure that data libraries are reproducible and
are used consistently in experimental applications.


\subsection{Verification and validation}

The reliability of physics data libraries rests on their rigorous and
comprehensive testing.
 
Effective development process models, based on iterative-incremental patterns,
embed verification testing in the development of a data library; nevertheless,
the still widespread adoption of the waterfall process model can be a source of
risk for the quality of the data released.

Evidence of major deficiencies in the verification process has been observed in 
a recent release of atomic data libraries \cite{tns_epics}. 
Substantial effort should be put on improving this critical area.


Validation of data libraries is a critical issue for their use in experimental applications.
Only a relatively small portion of widely used atomic data libraries has been
quantitatively validated with respect to experimental measurements.
Some validation results reported in the literature concern cross sections for
photon, electron and proton interactions, such as \cite{tns_rayleigh}, \cite{tns_beb}, \cite{tns_bebshell}, \cite{tns_photoel}, 
\cite{nss_compton}, 
\cite{arxiv_pair}, 
\cite{tns_pcross}, \cite{tns_pixe}, atomic relaxation (e.g.
\cite{tns_relax_prob}) and atomic parameters (e.g. \cite{tns_binding}).
Substantial gaps are still present in the validation of the
content of electromagnetic  data libraries, and limited funding is invested by major
organizations in this critical area, despite the extensive use of these physics instruments.

Major efforts are required to assess the reliability of data libraries, which in turn require
the availability of experimental measurements as a reference for validation tests.

Validation of both the electromagnetic and the nuclear data libraries involves a collaborative effort between
the providers of the data and the experimental community, which in turn benefits 
from the assessment of the reliability of the data.
Validation requires the availability of experimental measurements: while the
EXFOR \cite{EXFOR} database makes experimental nuclear data openly available and
the HEPData \cite{Maguire_2017} repository focuses on high energy data relevant
to Monte Carlo event generators \cite{buckley_2019}, no equivalent repository
exists yet for experimental atomic data, which currently must be manually
retrieved from the literature for validation tests of data libraries.

Many valuable experimental measurements are reported in the literature, but
they are not directly available in a digital format suitable to be used in
validation tests.
Their digitization requires substantial effort; despite the usefulness of these data,
no resources are invested for this purpose, apart from some individual efforts 
related to a few specific validation tests.

Hardly any funding is invested either into making new measurements specifically for the
purpose of validating the content of physics data libraries: in this respect, it
has been noted that computational science demands a paradigm shift regard to
validation experiments \cite{post_2005}.

Substantial progress is also needed on the epistemology of validation of physics
data libraries: statements of validity often rest on qualitative and subjective
appraisal rather than being grounded on sound statistical inference.

\section{The IDataLib project}

IDataLib is a scientific research project proposed by the authors of this paper to INFN
(Istituto Nazionale di Fisica Nucleare), an Italian research institution
dedicated to the study of the fundamental constituents of matter and the laws
that govern them.
It concerns physics data libraries, in
particular to address the critical situation of electromagnetic data libraries.
It  benefits from close collaboration with ENDF 
and with the United States' National Nuclear Data Center (NNDC), 
hosted by Brookhaven National Laboratory.
At the time of submitting this paper, the IDataLib identifier means
``Intended Data Library''; if the project is approved by the INFN management, it will stand for
``INFN Data Library''.
The project is articulated over three synergic areas:
\begin{itemize}
\item the electromagnetic data libraries themselves, namely their physical content,
\item the validation of the library content through comparison with experiment,
\item the access to experimental data collections for validation purposes.
\end{itemize}

\begin{figure}[htbp] 
\centering 
\includegraphics[width=.99\textwidth]{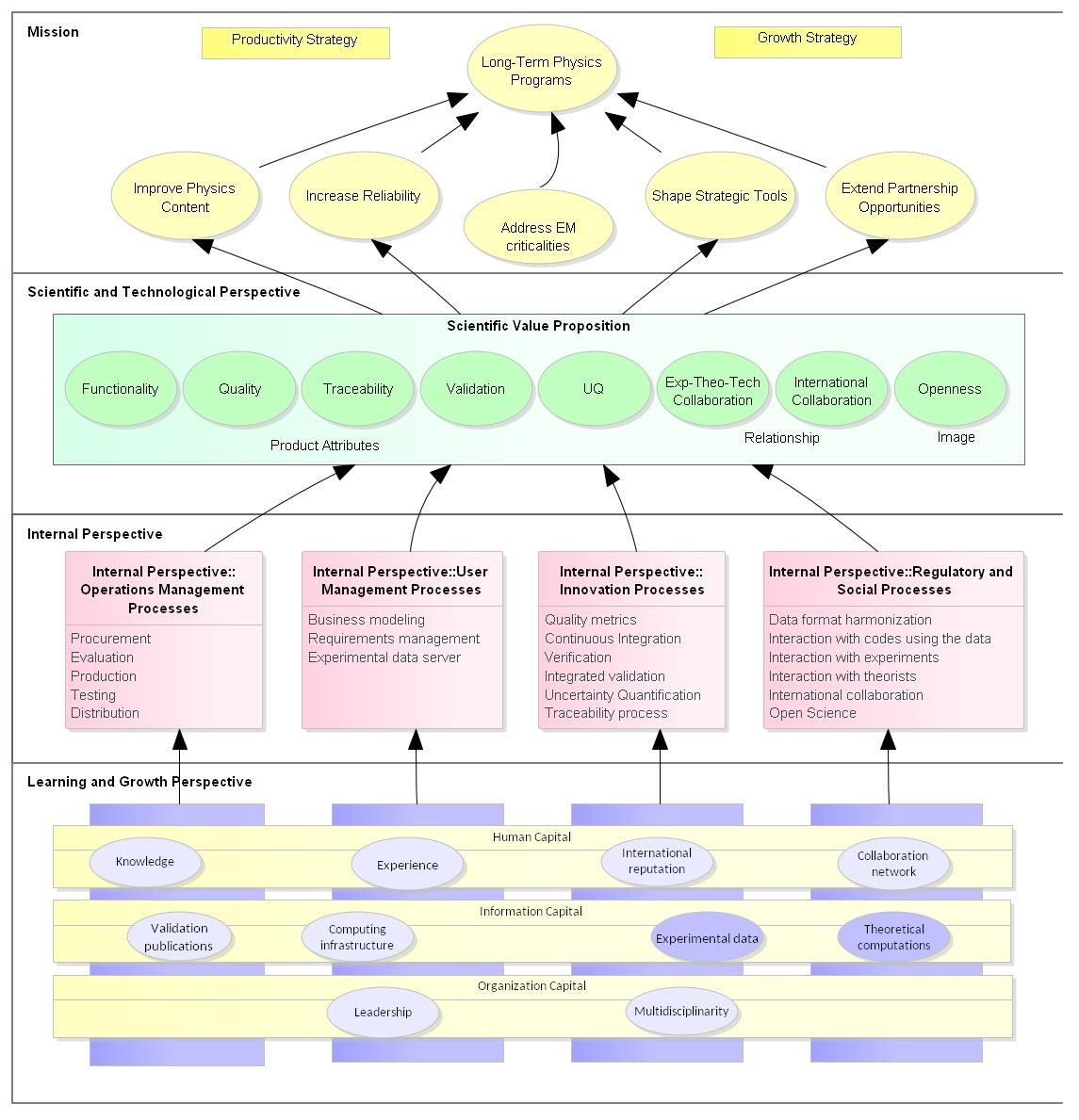} %
\caption{
\label{fig:strategy} 
Illustration of the strategy underlying the IDataLib project.}
\end{figure}

The strategy of the project is illustrated in figure~\ref{fig:strategy}.
It is supported  by an incremental and iterative process, which includes
monitoring the quality of the data and the associated software through objective, quantitative metrics.
All its activities are aligned with the spirit of  \textit{open science}.

Regarding the physics content, the IDataLib project intends to address the 
limitations of the current EEDL and EPDL. 
Previous validation tests have identified alternative calculation methods, which 
embody the state of the art for some cross sections better than the current tabulations.
These results provide the basis for improving the accuracy of the simulation of the 
physics processes they pertain to.
Further improvements and extensions will be investigated, taking into 
account the requirements of current and future experiments.

Special emphasis is devoted to the development of adequate tools for 
verification and validation testing, and to their use.

\begin{figure}
\centerline{\includegraphics[angle=0,width=8cm]{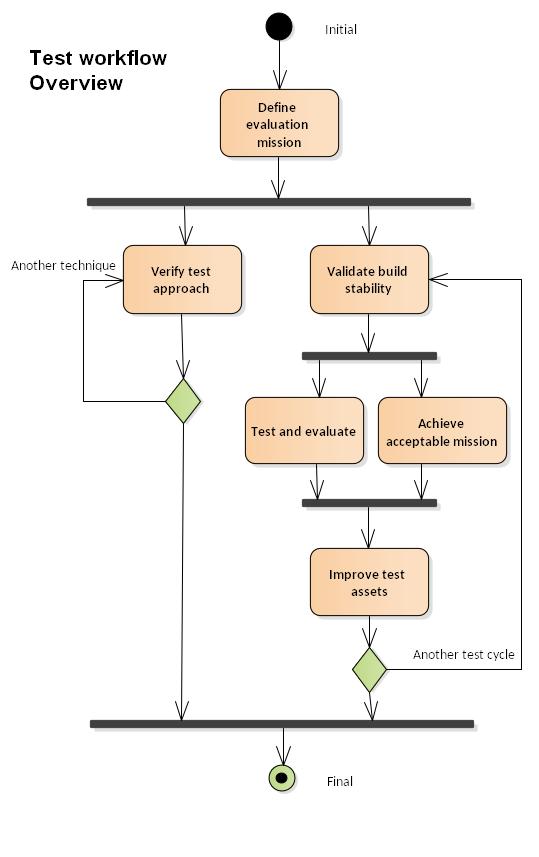}}
\caption{UML activity diagram, representing the workflow of the test process in the course of an iteration of the iterative-incremental 
process adopted in  the IDataLib project.}
\label{fig:test}
\end{figure}

The release of EPICS 2017 highlighted deficiencies in the verification of the 
data \cite{tns_epics} prior to their deployment.
The IDataLib project plans to define and apply a rigorous verification test process, supported by 
the development of appropriate testing tools.
The UML (Unified Modeling Language) activity diagram in figure \ref{fig:test}
illustrates the workflow of verification test.

The original validation process applied in several tests (e.g.
\cite{tns_rayleigh}, \cite{tns_beb}, \cite{tns_bebshell}, \cite{tns_photoel},
\cite{nss_compton}, \cite{arxiv_pair}, \cite{tns_pcross}, \cite{tns_pixe},
\cite{tns_relax_prob}, \cite{tns_binding}, \cite{tns_sandia1} \cite{tns_sandia2})
has gone into various refinements since its first development described in \cite{tns_sandia1}.
It consists of a two-stage process: the first stage encompasses goodness of fit
testing, while the second deals with the analysis of categorical data derived
from the outcome of the first stage. In both stages a variety of statistical
tests is used, which contributes to the robustness of the results by mitigating
the risk of possible systematic effects associated with peculiarities of the
mathematical formulation of the statistical tests.

Generalization of the related software into a validation test system is
planned in the context of the IDataLib project: it would be beneficial, as it
would facilitate its use in future validation tests, regardless of the specific
characteristics of the physics data and of the experimental measurements
involved in the validation.

At the time of writing this paper the approval of the IDataLib project  by 
the INFN management is still pending.

\section{Conclusion}

Physics data libraries are essential instruments in the experimental environment,
as they provide -- directly or indirectly, e.g. through their use in Monte Carlo particle transport codes --
the foundations to describe particle interactions with matter.
Detector design and optimization heavily rely on their key role in Monte Carlo simulation.

Extensive international attention is devoted to issues such as the openness,
transparency, traceability and reliability of physics data libraries used in
detector development and, in general, in the computational environment of
particle, nuclear, astroparticle and medical physics experimental research.
The elaboration of strategies to address open issues in these areas complements the continuing 
attention to the improvement and extension of the physics content of data libraries.


Collaboration among the various players in this field -- library developers, theoreticians,
experimental communities, mathematicians and statisticians -- is needed to address 
the complex needs of this domain.

The IDataLib project intends to address critical areas identified in 
the current situation of electromagnetic data libraries, which are a key instrument for 
experimental research in many fundamental and applied physics domains.

\acknowledgments

The authors are grateful to Sergio Bertolucci and David A. Brown for valuable discussions.


\bibliographystyle{JHEP}
\Urlmuskip=0mu plus 1mu\relax
\bibliography{bibliography_abbr}



\end{document}